\documentclass[10pt,twocolumn,twoside]{IEEEtran} 

\usepackage{cite}
\usepackage[dvips]{graphicx}

\usepackage[cmex10]{amsmath}
\usepackage{amssymb}

\DeclareMathOperator*{\argmin}{arg\,min}
\DeclareMathOperator*{\argmax}{arg\,max}

\newtheorem{theorem}{Theorem}
\newtheorem{corollary}{Corollary}
\newtheorem{lemma}{Lemma}

\newtheorem{remark}{Remark}
\newtheorem{definition}{Definition}

\begin{document}

\title{Online Recovery Guarantees and Analytical Results for {OMP}}
\author{Nazim Burak Karahanoglu,
        and~Hakan~Erdogan\\
\thanks{Nazim Burak Karahanoglu is  with the Advanced Technologies Research Institute, The Scientific and Technological Research Council of Turkey (TUBITAK), Kocaeli, Turkey. (email: karahanoglu@tubitak.gov.tr)}
\thanks{Hakan Erdogan is with the Faculty of Engineering and Natural Sciences, Sabanci University, Istanbul, Turkey. (email: haerdogan@sabanciuniv.edu)}
}
\maketitle

\begin{abstract}
Orthogonal Matching Pursuit (OMP) is a simple, yet empirically competitive algorithm for sparse recovery. Recent developments have shown that OMP guarantees exact recovery of $K$-sparse signals with $K$ or more than $K$ iterations if the observation matrix satisfies the restricted isometry property (RIP) with some conditions.
We develop RIP-based online guarantees for recovery of a $K$-sparse signal with more than $K$ OMP iterations.
Though these guarantees cannot be generalized to all sparse signals a priori, we show that they can still hold online when the state-of-the-art $K$-step recovery guarantees fail. In addition, we present bounds on the number of correct and false indices in the support estimate for the derived condition to be less restrictive than the $K$-step guarantees.
Under these bounds, this condition guarantees exact recovery of a $K$-sparse signal within $\frac{3}{2}K$ iterations, which is much less than the number of steps required for the state-of-the-art exact recovery guarantees with more than $K$ steps.
Moreover, we present phase transitions of OMP in comparison to basis pursuit and subspace pursuit, which are obtained after extensive recovery simulations involving different sparse signal types.
Finally, we empirically analyse the number of false indices in the support estimate, which indicates that these do not violate the developed upper bound in practice.

\end{abstract}

\section{Introduction}

%
%

Sparse recovery problem aims at finding the $K$-sparse signal $\mathbf{x}\in{\mathbb{R}}^{N}$ that satisfies a set of linear observations $\mathbf{y}\in{\mathbb{R}}^{M}$ where $K<M<N$.
Mathematically, this is expressed as
\begin{equation}
\mathbf{x}=\argmin\|\mathbf{x}\|_{0} \;\;\; s.t. \;\;\; \mathbf{y}=\mathbf{\Phi}\mathbf{x}, \nonumber
\end{equation}
where $\mathbf{\Phi}\in{\mathbb{R}}^{M\times{N}}$ is called the observation matrix, or the dictionary. Problems of this or similar forms appear for signal recovery or approximation in Compressed Sensing (CS) \cite{Donoho:CS, Tropp:CompMeth, Baraniuk:ModelCS, Maleki:TST, Karahanoglu:AOMPfull, FBP_DSP}, for finding sparse representations in overcomplete dictionaries \cite{Mallat:MP, Pati:OMP, Chen:BP}, etc.

Among others, Orthogonal Matching Pursuit (OMP) \cite{Pati:OMP} is a canonical greedy algorithm for sparse recovery. It aims at finding the support, i.e. the set of nonzero indices, of $\mathbf{x}$ one by one. At each iteration, OMP identifies the index corresponding to the column of $\mathbf{\Phi}$ which has maximum correlation to the residue of $\mathbf{y}$. Due to their simplicity and empirically competitive performance, OMP and its variants have been frequently used in sparse recovery and approximation problems \cite{Donoho:CS, Duarte:StructuredCS, Tropp:OMP, Shim:GOMP, Blumensath:GP, Blumensath:StWGP, Tropp:Greed}.

\subsection{Restricted Isometry Property}

Restricted isometry property (RIP) \cite{Candes:DecLP} has been acknowledged as an important means for obtaining theoretical guarantees in recovery and approximation problems. RIP is defined as follows:
\begin{definition}[Restricted Isometry Property]  A matrix $\mathbf{\Phi}$ is said to satisfy the $K$-RIP if there exists a Restricted Isometry Constant (RIC) $\delta_K$ satisfying $0<\delta_K<1$ and
\begin{equation}\label{Eq:RIP}
    (1-\delta_K)\|\mathbf{x}\|_2^2 \leq \|\mathbf{\Phi}\mathbf{x}\|_2^2 \leq (1+\delta_K)\|\mathbf{x}\|_2^2,
\end{equation}
\end{definition}
for all $\mathbf{x}$ where $\|\mathbf{x}\|_0 \leq K$. A matrix that satisfies RIP acts almost like an orthonormal system for sparse linear combinations of its columns \cite{Candes:DecLP}.
Random matrices with i.i.d. Gaussian or Bernoulli entries and matrices randomly selected from the discrete Fourier transform were shown to satisfy the RIP with high probabilities, when they satisfy some specific conditions on $K$, $M$ and $N$ \cite{Candes:NOptRec, Rudelson:SparseRec}.

RIP has been utilized for proving theoretical guarantees of exact recovery for many algorithms in the CS literature. These include convex relaxation \cite{Candes:DecLP, Candes:NOptRec, Candes:RIP} and greedy algorithms such as Regularized OMP (ROMP) \cite{Needell:ROMP_IEEE}, Compressive Sampling Matching Pursuit (CoSaMP) \cite{Needell:CoSAMP}, Subspace Pursuit (SP) \cite{Dai:SP}, Iterative Hard Thresholding (IHT) \cite{Blumensath:IHT2}, etc.

\subsection{Recent Developments in Theoretical Analysis of OMP}
\label{sec:RecentOMP}

Initial contributions on the theoretical analysis of OMP have concentrated on coherence \cite{Tropp:Greed} or probability analysis \cite{Tropp:OMP, Fletcher:OMP}.
Recently, Davenport and Wakin have presented a straightforward $K$-step analysis of OMP based on RIP \cite{Davenport:AnalysisOMP}. Their work states that OMP guarantees exact recovery of any $K$-sparse signal from noise-free measurements in $K$ iterations if $\mathbf{\Phi}$ fulfills RIP with RIC satisfying $\delta_{K+1}<\frac{1}{3\sqrt{K}}$.
Lately, Wang and Shim have proven a less restricted bound for OMP \cite{Wang:OMP_analysis} which we visit in the following theorem:

\begin{theorem}[Exact recovery condition for OMP \cite{Wang:OMP_analysis}]
\label{Thrm_Wang}
OMP perfectly recovers any $K$-sparse signal from noise-free measurements in $K$ iterations if the observation matrix $\mathbf{\Phi}$ satisfies RIP with
\begin{equation}\label{Wang:OMP_result}
\delta_{K+1} < \frac{1}{\sqrt{K}+1}.
\end{equation}
\end{theorem}
Note that Theorem~\ref{Thrm_Wang} represents a special case of Theorem~\ref{Thrm_OMP1}, which is introduced below. According to the $K$-step recovery analyses in \cite{Davenport:AnalysisOMP} and \cite{Wang:OMP_analysis}, OMP requires $M=O(K^2\log(N))$ measurements for exact recovery in $K$ iterations.

Due to the intuitive improvements in the OMP recovery accuracy with more than $K$ iterations, theoretical analyses have been performed for this case as well.
\cite{Zhang:OMP_RIP} states that OMP exactly recovers all $K$-sparse signals within $30K$ iterations when $\mathbf{\Phi}$ satisfies RIP with $\delta_{31K}\leq\frac{1}{3}$.
The number of necessary iterations has been first reduced to $12K$ with $\delta_{22K}\leq\frac{1}{6}$ \cite{Foucart:WOMP}, and later to $6K$ with $\delta_{\lfloor8.93K\rfloor}<0.03248$ \cite{Wang:Imp_OMP}.
According to these findings, OMP necessitates $O(K\log(N))$ measurements for exact recovery with more than $K$ iterations, an improvement over the $O(K^2\log(N))$ measurements required by the $K$-step guarantees of \cite{Davenport:AnalysisOMP} and \cite{Wang:OMP_analysis}.
However, these more than $K$-step guarantees necessitate $6K$ to $30K$ iterations, which is mostly beyond the practical limits in many applications.
That is, for many applications, these state-of-the-art more than $K$-step guarantees are not useful anymore.

\subsection{Our Contributions}

In this manuscript, we aim at providing an online recovery analysis of the OMP algorithm. For this purpose, we extend the theoretical analysis in \cite{Wang:OMP_analysis} to cover for more than $K$ iterations. In addition, we demonstrate OMP recovery with both $K$ and more than $K$ iterations via phase transitions in comparison to some other mainstream recovery algorithms. We concentrate on the residue-based termination rule, which terminates when the residue of the observed vector gets small enough, in contrast to the sparsity-based termination, which limits the number of iterations by $K$. To avoid ambiguity, we use the term OMP$_K$ to indicate the sparsity-based termination rule, and OMP$_e$ for the residue-based termination.

As for the theoretical analyses, we develop a model by extending the findings of \cite{Wang:OMP_analysis}
to cover more than $K$ iterations in Section~\ref{Sec:Analysis}.
In Theorem~\ref{Thrm_Main}, we derive RIP-based online guarantees for the success of an OMP$_e$ iteration.
Next, we present online recovery guarantees for OMP$_e$ in Theorem~\ref{Thrm_OMP1}, which is obtained by generalizing Theorem~\ref{Thrm_Main} for all consequent iterations.
Since both Theorem~\ref{Thrm_Main} and Theorem~\ref{Thrm_OMP1} depend on the number of correct and false indices in a particular support estimate, generalization of these results for all $K$-sparse signals necessitates assuring the existence of support estimates with sufficiently large number of correct detections.
Unfortunately, we cannot provide such guarantees.
However, OMP$_e$ obviously enjoys all theoretical guarantees of OMP$_K$ for the noise-free case\footnote{It is obvious that the first $K$ steps of both variants are identical. In parallel, Theorem~\ref{Thrm_Wang} is a special case of Theorem~\ref{Thrm_OMP1}. This theoretically guarantees this intuitive fact.}.
Furthermore, Section~\ref{Sec:CompOMPK_e}, which deals with the validity of the developed online guarantees in practice, states that Theorem~\ref{Thrm_OMP1} becomes less restrictive than Theorem~\ref{Thrm_Wang} when the number of correct and false detections in the support estimate satisfy some conditions.
Under these conditions, it becomes possible to satisfy Theorem~\ref{Thrm_OMP1} although Theorem~\ref{Thrm_Wang} fails.
If satisfied under these conditions, Theorem~\ref{Thrm_OMP1} provides online exact recovery guarantees for a $K$-sparse signal within $\frac{3}{2}K$ iterations.
This number is clearly less than the $6K$ to $30K$ iterations, which are necessary for the state-of-the-art exact recovery guarantees of \cite{Zhang:OMP_RIP}, \cite{Foucart:WOMP}, and \cite{Wang:Imp_OMP}.

Finally, we present empirical phase transition curves for three different types of sparse signals in order to demonstrate the  recovery performance of OMP in comparison to some other mainstream algorithms. To the best of our knowledge, phase transitions comparing the two OMP versions with BP and SP for different coefficient distributions appears for the first time. Hence, they are not only important for revealing the actual difference between the two OMP variants but also comparing OMP with two of the other well-established, and mostly more credited algorithms. In addition, we provide histograms of the number of false indices after successful OMP$_e$ termination in Section~\ref{Sec:Results_Hist}. This demonstrate that the upper bound on the number of false indices which the online guarantees require is loose in practice.

\subsection{Notation}

Let us now define the notation we use throughout this manuscript. $T$ denotes the correct support of $\mathbf{x}$. $T^l = \{t_1,t_2,...,t_l\}$ is the support estimate for $\mathbf{x}$ after the $l$th iteration of OMP, where $t_i$ is the index selected at the $i$th iteration. $n_c$ and $n_f$ denote the number of correct and false indices in ${T^l}$, respectively, i.e. $|T \cap T^l| = n_c$ and $|T^l - T| = n_f$. The observation matrix is decomposed as $\mathbf{\Phi} = [\mathbf{\phi}_1 \mathbf{\phi}_2 ... \mathbf{\phi}_N]$, where $\mathbf{\phi}_i$ is the $i$th column vector of $\mathbf{\Phi}$. $\mathbf{\Phi}_{T}$ denotes the matrix consisting of the columns of $\mathbf{\Phi}$ indexed by $T$, and $\mathbf{x}_{T}$ is the vector consisting of the elements of $\mathbf{x}$ indexed by $T$. $\mathbf{r}^l$ is the residue after the orthogonal projection of $\mathbf{y}$ onto $\mathbf{\Phi}_{T^l}$ by the end of the $l$th iteration. Finally, $\mathbf{\Phi}^*$ denotes the conjugate of a matrix $\mathbf{\Phi}$.

\section{Theoretical Analysis} \label{Sec:Analysis}

\subsection{Preliminaries}

The analysis we present in the next section is based on a number of preliminary results, which are discussed below. These include some observations which are well-known in the CS community as well as some results which we derive in this manuscript for our purposes.
Specifically, Lemma~\ref{lemma_SP1} is a direct consequence of RIP, while Lemma~\ref{lemma_SP2} and Corollary~\ref{Cor_ExRec1} are from \cite{Dai:SP} and \cite{Needell:CoSAMP}, respectively. 
Lemma~\ref{Lemma_ExRec2} is simply derived from Corollary~\ref{Cor_ExRec1}, and Remark~\ref{Cor_ExRec3} is a direct consequence of Lemma~\ref{Lemma_ExRec2}. Finally, we derive Lemma~\ref{lemma_ExRec1}, which we will later exploit for comparing the RIP bound of Theorem~\ref{Thrm_Wang} with our result. The proofs are omitted either if they are very trivial, or they are already present in the corresponding references.

\begin{lemma}[Direct Consequence of RIP]
Let $I \subset \{1,2,...,N\}$. For any arbitrary vector $\mathbf{z} \in \mathbb{R}^{|I|}$
\begin{equation}
(1-\delta_{|I|})\|\mathbf{z}\|_2 \leq \|\mathbf{\Phi}_I^*\mathbf{\Phi}_I\mathbf{z}\|_2 \leq (1+\delta_{|I|})\|\mathbf{z}\|_2. \nonumber
\end{equation}
\label{lemma_SP1}
\end{lemma}

\begin{lemma}[Lemma 1 in \cite{Dai:SP}] Let $I,J \subset \{1,2,...,N\}$ such that $I \cap J = \emptyset$. For any arbitrary vector $\mathbf{z} \in \mathbb{R}^{|J|}$
\label{lemma_SP2}
\begin{equation}
\|\mathbf{\Phi}_I^*\mathbf{\Phi}_J\mathbf{z}\|_2 \leq \delta_{|I|+|J|}\|\mathbf{z}\|_2. \nonumber
\end{equation}
\end{lemma}

\begin{corollary}[Corollary 3.4 in \cite{Needell:CoSAMP}] For every positive integer c and r
\label{Cor_ExRec1}
\begin{equation} \label{Corr3.4}
\delta_{cr}<c\delta_{2r}.
\end{equation}
\end{corollary}

\begin{lemma}
\label{Lemma_ExRec2}
For any positive integer $K$
\begin{equation} 
\delta_{K+1} > \frac{\delta_{3\lceil K/2 \rceil}}{3},
\end{equation}
where $\lceil z \rceil$ denotes the ceiling of $z$, i.e. the smallest integer greater than or equal to $z$.
\end{lemma}

\begin{IEEEproof}
Lemma~\ref{Lemma_ExRec2} is a consequence of Corollary~\ref{Cor_ExRec1}. We first replace $c=3$ and $r = \lceil K/2 \rceil$ into (\ref{Corr3.4}). By rearranging terms, we get
\begin{equation}
  \delta_{2\lceil K/2 \rceil} > \frac{\delta_{3\lceil K/2 \rceil}}{3}. \nonumber
\end{equation}
$K+1 \geq 2\lceil K/2 \rceil$ hold by definition. Following the monotonicity of RIC, we have $\delta_{K+1} \geq \delta_{2\lceil K/2 \rceil}$. Hence, we can write
\begin{IEEEeqnarray}[]{rcl}
  \delta_{K+1} &{}\geq{}& \delta_{2\lceil K/2 \rceil} \nonumber \\
                &{}>{}& \frac{\delta_{3\lceil K/2 \rceil}}{3}. \nonumber
\end{IEEEeqnarray}
\end{IEEEproof}

\begin{remark}[Direct consequence of Lemma~\ref{Lemma_ExRec2}]
\label{Cor_ExRec3}
Theorem~\ref{Thrm_Wang} is violated if
\begin{equation} \label{Eq_Cor_ExRec3}
\delta_{3\lceil K/2 \rceil} \geq \frac{3}{\sqrt{K}+1}.
\end{equation}
\end{remark}

\begin{IEEEproof} According to Lemma~\ref{Lemma_ExRec2}, it is clear that (\ref{Eq_Cor_ExRec3}) contradicts Theorem~\ref{Thrm_Wang}.
\end{IEEEproof}


\begin{lemma}
\label{lemma_ExRec1}
Assume $K \geq 25$. There exists at least one positive integer $n_c<K$ that satisfies
\begin{equation}
\label{Eq:Lemma_ExRec1_1}
 \frac{3}{\sqrt{K}+1} \leq \frac{1}{\sqrt{K-n_c}+1}.
\end{equation}
Moreover, such values of $n_c$ are bounded by
\begin{equation}
\label{Eq:n_cBound}
 K > n_c \geq \frac{8K+4\sqrt{K}-4}{9}.
\end{equation}
\end{lemma}
\begin{IEEEproof}
Set $K-n_c = sK$ where $0<s<1$. 
Replacing $s$ into (\ref{Eq:Lemma_ExRec1_1}), we get
\begin{equation}
 \frac{3}{\sqrt{K}+1} \leq \frac{1}{\sqrt{sK}+1}. \nonumber
\end{equation}
Arranging the terms, we obtain the following bound for $s$:
\begin{equation}
 s \leq \left( \frac{\sqrt{K}-2}{3\sqrt{K}} \right)^2. \nonumber
\end{equation}
Then, the lower bound for $n_c$ is obtained as
\begin{IEEEeqnarray}[]{rcl}
n_c &{}={}& (1-s)K \nonumber \\
    &{}\geq{}& \frac{8K+4\sqrt{K}-4}{9}.
    \label{Eq:Lemma_ExRec1_2}
\end{IEEEeqnarray}

On the other hand, $n_c<K$ requires $sK = K-n_c \geq 1$. Hence, $K$ should satisfy
\begin{IEEEeqnarray}[]{rcl}
 K &{}\geq{}& \frac{1}{s} \nonumber \\
   &{}\geq{}& \left(\frac{3\sqrt{K}}{\sqrt{K}-2}\right)^2. \nonumber
\end{IEEEeqnarray}
Rearranging terms we get
\begin{equation}
 K \geq 5\sqrt{K}, \nonumber
  \label{Eq:Lemma_ExRec1_3}
\end{equation}
which is satisfied when $K \geq 25$. Combining this with (\ref{Eq:Lemma_ExRec1_2}), we conclude (\ref{Eq:Lemma_ExRec1_1}) is satisfied if
\begin{equation}
K > n_c \geq \frac{8K+4\sqrt{K}-4}{9} \nonumber
\end{equation}
for $K\geq25$.

\end{IEEEproof}

\subsection{Success of a Single Iteration of OMP$_e$} \label{Sec:SuccessSingIter}

Having presented the necessary preliminary results, we can now move on to the analysis of OMP$_e$. We start with success of a single iteration, for which the theorem below states a sufficient condition depending on the number of correct and false indices in the support estimate.

\begin{theorem}
\label{Thrm_Main}
Let $|{T^l \cap T}| = n_c$ and $|{T^l - T}| = n_f$ after iteration $l$. Then iteration $l+1$ will be successful, i.e. $t_{l+1} \in T - T^l$ if $\mathbf{\Phi}$ satisfies RIP with
\begin{equation}
\delta_{K+n_f+1} < \frac{1}{\sqrt{K-n_c}+1}.
\label{AOMP_res1}
\end{equation}
\end{theorem}
\begin{IEEEproof}
As $\mathbf{r}^l$ is the projection error of $\mathbf{y}$ onto $ \mathbf{\Phi}_{T^l}$, we have ${\mathbf{r}^l \perp \mathbf{\Phi}_{T^l}}$. Therefore, $\langle\mathbf{\phi}_i,\mathbf{r}^l\rangle = 0$ for all $i \in T^l$. We can then write
\begin{IEEEeqnarray}[]{rcl}
\label{Corr_TTl}
\|\mathbf{\Phi}_{T\cup T^l}^*\mathbf{r}^l\|_2^2 &{}={}& \sum_{i \in T\cup T^l}\langle\mathbf{\phi}_i,\mathbf{r}^l\rangle^2 \nonumber \\
&{}={}& \sum_{i \in T - T^l}\langle\mathbf{\phi}_i,\mathbf{r}^l\rangle^2,
\end{IEEEeqnarray}
where the righthand side of (\ref{Corr_TTl}) contains only ${K - n_c}$ nonzero terms. Combining (\ref{Corr_TTl}) and the norm inequality, we obtain
\begin{equation} \label{Eq_infNormBound}
\|\mathbf{\Phi}_{T\cup T^l}^*\mathbf{r}^l\|_\infty \geq \frac{1}{\sqrt{K-n_c}}\|\mathbf{\Phi}_{T\cup T^l}^*\mathbf{r}^l\|_2.
\end{equation}
Now, let $\tilde{\mathbf{x}}$ denote the estimate of $\mathbf{x}$ after iteration $l$. Then, $\mathbf{r}^l$ can be written as
\begin{IEEEeqnarray}[]{rcl}\label{residue}
\mathbf{r}^l &{}={}& y - \mathbf{\Phi}_{T^l}\tilde{\mathbf{x}}_{T^l} \nonumber \\
             &{}={}& \mathbf{\Phi}_T\mathbf{x}_T - \mathbf{\Phi}_{T^l}\tilde{\mathbf{x}}_{T^l} \nonumber \\
             &{}={}& \mathbf{\Phi}_{T\cup T^l}\mathbf{z}, \nonumber
\end{IEEEeqnarray}
where $\mathbf{z}$ is a vector of length $K+n_f$. By Lemma~\ref{lemma_SP1}, we obtain
\begin{IEEEeqnarray}[]{rcl} \label{Eq_CorrBound}
\|\mathbf{\Phi}_{T\cup T^l}^* \mathbf{r}^l\|_2 &{}={}& \|\mathbf{\Phi}_{T\cup T^l}^*\mathbf{\Phi}_{T\cup T^l}\mathbf{z}\|_2 \nonumber \\
&{}\geq{}& (1-\delta_{K+n_f}) \|\mathbf{z}\|_2.
\end{IEEEeqnarray}
Replacing (\ref{Eq_CorrBound}) into (\ref{Eq_infNormBound}) yields
\begin{equation} \label{Eq_infNormBound2}
\|\mathbf{\Phi}_{T\cup T^l}^*\mathbf{r}^l\|_\infty \geq \frac{1-\delta_{K+n_f}}{\sqrt{K-n_c}}\|\mathbf{z}\|_2 .
\end{equation}
The selection rule for the index $t_{l+1}$ at iteration $l+1$ is defined as
  \begin{equation}\label{t_l+1}
  t_{l+1} = \argmax\limits_{i} \left| \langle \mathbf{\phi}_i,\mathbf{r}^l \rangle \right|.
  \end{equation}
Combining this definition with (\ref{Eq_infNormBound2}), we obtain
\begin{IEEEeqnarray}[]{rcl}
| \langle \mathbf{\phi}_{t_{l+1}},\mathbf{r}^l \rangle | &{}={}& \| \mathbf{\Phi}^*\mathbf{r}^l \|_\infty \nonumber \\
                                                &{}\geq{}& \|\mathbf{\Phi}_{T\cup T^l}^*\mathbf{r}^l\|_\infty \nonumber \\
&{}\geq{}& \frac{1-\delta_{K+n_f}}{\sqrt{K-n_c}}\|\mathbf{z}\|_2. \nonumber
\end{IEEEeqnarray}

Now, suppose that iteration $l+1$ fails, i.e. $t_{l+1} \notin T\cup T^l$. Then, we can write
\begin{IEEEeqnarray}[]{rcl} 
| \langle \mathbf{\phi}_{t_{l+1}},\mathbf{r}^l \rangle | &{}={}& \|\mathbf{\phi}_{t_{l+1}}^*\mathbf{\Phi}_{T\cup T^l}\mathbf{z}\|_2 \nonumber \\
&{}\leq{}&\delta_{K+n_f+1}\|\mathbf{z}\|_2  \nonumber
\end{IEEEeqnarray}
by Lemma~\ref{lemma_SP2}. Clearly, this never occurs if
\begin{equation}
 \frac{1-\delta_{K+n_f}}{\sqrt{K-n_c}}\|\mathbf{z}\|_2 > \delta_{K+n_f+1}\|\mathbf{z}\|_2 \nonumber
\end{equation}
or equivalently
\begin{equation} \label{Cond1}
 \sqrt{K-n_c}\;\delta_{K+n_f+1} + \delta_{K+n_f} < 1
\end{equation}
Following the monotonicity of RIC, we know that $\delta_{K+n_f+1} \geq \delta_{K+n_f}$. Hence, (\ref{Cond1}) is guaranteed when
\begin{equation}\label{delta_bound1}
 \sqrt{K-n_c}\;\delta_{K+n_f+1} + \delta_{K+n_f+1} < 1, \nonumber
\end{equation}
which is equivalent to
\begin{equation}
\delta_{K+n_f+1} < \frac{1}{\sqrt{K-n_c}+1}.
\label{final_theorem}
\end{equation}
Hence, $t_{l+1} \in T\cup T^l$ when (\ref{final_theorem}) holds.  We also know that $\langle\mathbf{\phi}_i,\mathbf{r}^l\rangle = 0$ for all $i \in T^l$. Therefore, a selected index cannot be selected again in the following iterations, i.e. $t_{l+1} \notin T^l$. In combination with (\ref{final_theorem}) this directly leads to $t_{l+1} \in T - T^l$, that is iteration $l+1$ will be successful.
\end{IEEEproof}

Theorem~\ref{Thrm_Main} and Theorem~\ref{Thrm_Wang} are naturally related. Theorem~\ref{Thrm_Wang} is based on the fact that the RIP condition in (\ref{Wang:OMP_result}) guarantees exact recovery of an iteration, provided that all previous iterations have been successful\footnote{Note that the success condition of an OMP$_K$ iteration corresponds to the case $n_f=0$ in (\ref{AOMP_res1}). The proof of Theorem~\ref{Thrm_Wang} presented in \cite{Wang:OMP_analysis} is based on this restricted condition.}.
The dependency on the success of all previous iterations is necessary for exact recovery in $K$ iterations.
In contrast, Theorem~\ref{Thrm_Main} removes the dependency of the success condition on the success of all previous iterations\, generalizing the success condition of a single iteration to a broader extend which can handle failures among previous iterations.
However, as a trade-off, we end up with an online guarantee that depends on the number of correct and incorrect indices in the support estimate of a specific iteration.

\subsection{Online Recovery Guarantees for OMP$_e$} \label{Sec:OMPeAnalysis}

Online recovery guarantees for OMP$_e$ can be obtained by generalization of Theorem~\ref{Thrm_Main} to all the following iterations until the successful termination of the algorithm. That is, the conditions in Theorem~\ref{Thrm_Main} do guarantee the success of not only a particular iteration, but also all the following ones. This is stated in the following theorem:
\begin{theorem}
\label{Thrm_OMP1}
Let $|{T^l \cap T}| = n_c$ and $|{T^l - T}| = n_f$ after iteration $l$. Then, OMP$_e$ perfectly recovers a $K$-sparse signal in a total of $K+n_f$ iterations if $\mathbf{\Phi}$ satisfies RIP with
  \begin{equation}\label{OMP_guarantee1}
  \delta_{K+n_f+1} < \frac{1}{\sqrt{K-n_c}+1}.
  \end{equation}
\end{theorem}
\begin{IEEEproof}
We prove Theorem~\ref{Thrm_OMP1} by induction.
According to Theorem~\ref{Thrm_Main}, (\ref{OMP_guarantee1}) already guarantees success of the iteration $l+1$.
As a result of this, $t_{l+1} \in T - T^l $ and $T_{l+1}$ will contain $n_c+1$ correct indices.
Since the right hand side of (\ref{OMP_guarantee1}) increases monotonically with the number of correct indices in the support estimate, iteration $l+2$ requires a less restrictive RIP constraint than iteration $l+1$ does.
Therefore, (\ref{OMP_guarantee1}) also guarantees also the success of the iteration $l+2$ in addition to the iteration $l+1$.
By induction, this applies to all of the following iterations as each of them requires a less restrictive RIP condition.
Hence, after $K-n_c$ additional iterations, i.e. after the iteration $K+n_f$, the support estimate $T_{K+n_f}$ will contain $K$ correct indices, i.e. $T \subset T_{K+n_f}$.
Finally (\ref{OMP_guarantee1}) guarantees that the orthogonal projection coefficients of $\mathbf{y}$ onto $T_{K+n_f}$ yield exactly $\mathbf{x}$.
\end{IEEEproof}

Being an extension of Theorem~\ref{Thrm_Main}, Theorem~\ref{Thrm_OMP1} also depends on $n_c$ and $n_f$.
This allows online recovery guarantees which cover more than $K$ iterations.
Yet, this also prevents us from generalizing our results as exact recovery guarantees for all $K$-sparse signals since the existence of intermediate steps with enough number of correct indices in addition to a small number of false indices is hard to guarantee.
We cannot provide a proof of this for the time being, leaving it as a future work.
However, we investigate the possibility of the existence of such support estimates for some particular conditions in the next section. In addition, we also would like to refer the reader to Section~\ref{Sec:Results_Hist}, where we investigate the number of incorrect indices empirically by histograms. These histograms demonstrate that $n_f$ is indeed bounded in practice.

\subsection{On the Validity of the Online Guarantees}
\label{Sec:CompOMPK_e}


In order for the online recovery condition in Theorem~\ref{Thrm_OMP1} to be meaningful, it should also be shown that this condition can be satisfied online at some intermediate iteration in case the $K$-step recovery condition of Theorem~\ref{Thrm_Wang} fails.
For this purpose, we provide below a comparison of the RIP conditions in Theorem~\ref{Thrm_OMP1} and Theorem~\ref{Thrm_Wang}. This comparison proves that Theorem~\ref{Thrm_OMP1} requires a less restrictive bound on the RIC than Theorem~\ref{Thrm_Wang} does when $n_c$ and $n_f$ are large and small enough, respectively.

In order to state that (\ref{OMP_guarantee1}) implies a less restrictive condition than (\ref{Wang:OMP_result}) at least for some particular cases, we need to compare the two bounds:
\begin{equation}
\delta_{K+1} < \frac{1}{\sqrt{K}+1}\;\; \longleftrightarrow\;\; \delta_{K+n_f+1} < \frac{1}{\sqrt{K-n_c}+1} \nonumber
\end{equation}
Unfortunately, that right and left-hand sides of the two bounds are constraints in the same direction:
\begin{IEEEeqnarray}[]{rcl}
\delta_{K+n_f+1} &{}\geq{}& \delta_{K+1}, \nonumber \\
\frac{1}{\sqrt{K-n_c}+1} &{}\geq{}& \frac{1}{\sqrt{K}+1}.  \nonumber
\end{IEEEeqnarray}
Hence, it is not possible to compare these two conditions directly.
Intuitively, when $n_f$ is small, and $n_c$ is large, we expect Theorem~\ref{Thrm_OMP1} to be less restrictive.
To illustrate, consider $n_f = 1$ and $n_c \gg n_f$.
In this case, Theorem~\ref{Thrm_OMP1} requires an RIP condition based on $\delta_{K+2}$ instead of $\delta_{K+1}$ of Theorem~\ref{Thrm_Wang}, i.e. two RIC's are practically very close to each other.
However, the upper bound in (\ref{OMP_guarantee1}) is significantly larger than the one in (\ref{Wang:OMP_result}) because of $n_c$ being large. Hence, (\ref{OMP_guarantee1}) becomes practically less restrictive in this situation.

Despite the intuitive  reasoning, exact mathematical comparison of these two conditions is tricky since it is not easy to obtain a tight bound on $\frac{\delta_{K+n_f+1}}{\delta_{K+1}}$ for all $n_f$.
However, even by employing a loose bound on $\delta_{K+n_f+1} / \delta_{K+1}$, we can show that (\ref{OMP_guarantee1}) becomes less restrictive than (\ref{Wang:OMP_result}) for some particular cases:
\begin{theorem}
\label{Thrm_OMP2}
Assume that $K\geq25$, $1 \leq n_f < \lceil K/2 \rceil$ and $n_c$ satisfies
\begin{equation}
\label{Eq:n_cBound2}
 K > n_c \geq \frac{8K+4\sqrt{K}-4}{9}
\end{equation}
at iteration $l$. Then, (\ref{OMP_guarantee1}) becomes less restrictive than (\ref{Wang:OMP_result}) at iteration $l$.
In such a case, the online recovery guarantees of Theorem~\ref{Thrm_OMP1} might be satisfied, even though $K$-step recovery cannot be guaranteed. Moreover, if Theorem~\ref{Thrm_OMP1} is satisfied under these conditions, OMP$_e$ is guaranteed to provide exact recovery within $\frac{3}{2}K$ iterations.
\end{theorem}
\begin{IEEEproof} Assume that
\begin{equation}
\label{Eq_Thrm_OMP2_1}
\delta_{K+n_f+1} \geq \frac{3}{\sqrt{K}+1}.
\end{equation}
Since $n_f < \left\lceil \frac{K}{2} \right\rceil$, we observe that $3\left\lceil \frac{K}{2} \right\rceil \geq K+n_f+1$. Following the monotonicity of RIC, we obtain
\begin{equation}
\label{Eq_Thrm_OMP2_2}
\delta _{3\left\lceil \frac{K}{2} \right\rceil} \geq \frac{3}{\sqrt{K}+1}.
\end{equation}
Remark~\ref{Cor_ExRec3} guarantees failure of (\ref{Wang:OMP_result}) for this case\footnote{This accomplies with the OMP$_K$ failure following the assumption $n_f \geq 1$.}.

On the other hand, Lemma~\ref{lemma_ExRec1} leads to
\begin{equation}
 \frac{3}{\sqrt{K}+1} \leq \frac{1}{\sqrt{K-n_c}+1} \nonumber
\end{equation}
when (\ref{Eq:n_cBound2}) is satisfied and $K\geq25$. Hence, there exists some $\delta_{K+n_f+1}$ such that
\begin{equation}
\label{Eq_OMP_1}
 \frac{3}{\sqrt{K}+1} \leq \delta_{K+n_f+1} \leq \frac{1}{\sqrt{K-n_c}+1}. \nonumber
\end{equation}
Clearly, $\delta_{K+n_f+1}$ values in this range satisfy (\ref{OMP_guarantee1}).

To conclude, when the parameters $K$, $n_f$ and $n_c$ satisfy the assumptions, there exists some $\delta_{K+n_f+1}$ which fulfill (\ref{OMP_guarantee1}), though (\ref{Wang:OMP_result}) does not hold for $\delta_{K+1}$.
Then, (\ref{OMP_guarantee1}) becomes less restrictive than (\ref{Wang:OMP_result}), and the online recovery guarantees of Theorem~\ref{Thrm_OMP1} might still be satisfied even though $K$-step recovery cannot be guaranteed for this range of parameters. In such a case, Theorem~\ref{Thrm_OMP1} guarantees exact recovery within $\frac{3}{2}K$ iterations since $n_f < \left\lceil\frac{K}{2}\right\rceil$ and all the following iterations are guaranteed to be successful.
\end{IEEEproof}

Theorem~\ref{Thrm_OMP2} states one particular case where the online guarantees of Theorem~\ref{Thrm_OMP1} turn into a less restrictive condition than the $K$-step exact recovery guarantees. Although we cannot yet generalize them, the presented online recovery guarantees can explain recovery of at least some particular sparse instances by OMP$_e$ in practice. Moreover, when the conditions of Theorem~\ref{Thrm_OMP2} are satisfied, exact recovery is possible within $\frac{3}{2}K$ iterations. This number is clearly much less than the $6K$ iterations which are needed for exact recovery of all $K$-sparse signals with OMP$_e$.

Note that the assumptions $K \geq 25$ and (\ref{Eq:n_cBound2}) in Theorem~\ref{Thrm_OMP2} rely on $n_f < \left\lceil \frac{K}{2} \right\rceil$.
This upper bound is chosen specifically in order to be able to establish (\ref{Eq_Thrm_OMP2_2}).
In other words, both $K\geq25$ and (\ref{Eq:n_cBound2}) actually apply for the boundary condition $n_f = \left\lceil \frac{K}{2} \right\rceil-1$.
These conditions are necessary to prove Theorem~\ref{Thrm_OMP2}.
However, we believe that these bounds are loose.
We intuitively expect that Theorem~\ref{Thrm_OMP2} also holds for smaller lower bounds on $K$ and $n_c$.
That is, the online recovery guarantees are expected to turn into less restrictive conditions for smaller $K$ and $n_c$ values as well.
Moreover, these bounds may be further improved with a tighter upper bound on $n_f$.
These issues may be addressed theoretically by tighter upper bounds on $\frac{\delta_{K+n_f+1}}{\delta_{K+1}}$ as future work.
Nonetheless, we analyse $n_f$ for successful OMP$_e$ recoveries via histograms in the next section. These indicate that the bound $n_f < \left\lceil \frac{K}{2} \right\rceil$ is usually loose in practice.

\section{Empirical Analysis} \label{Sec:Results}

\subsection{Phase Transitions}  \label{Sec:Results_PT}

\begin{figure*}[!htb]
\begin{center}
\includegraphics[width=\linewidth]{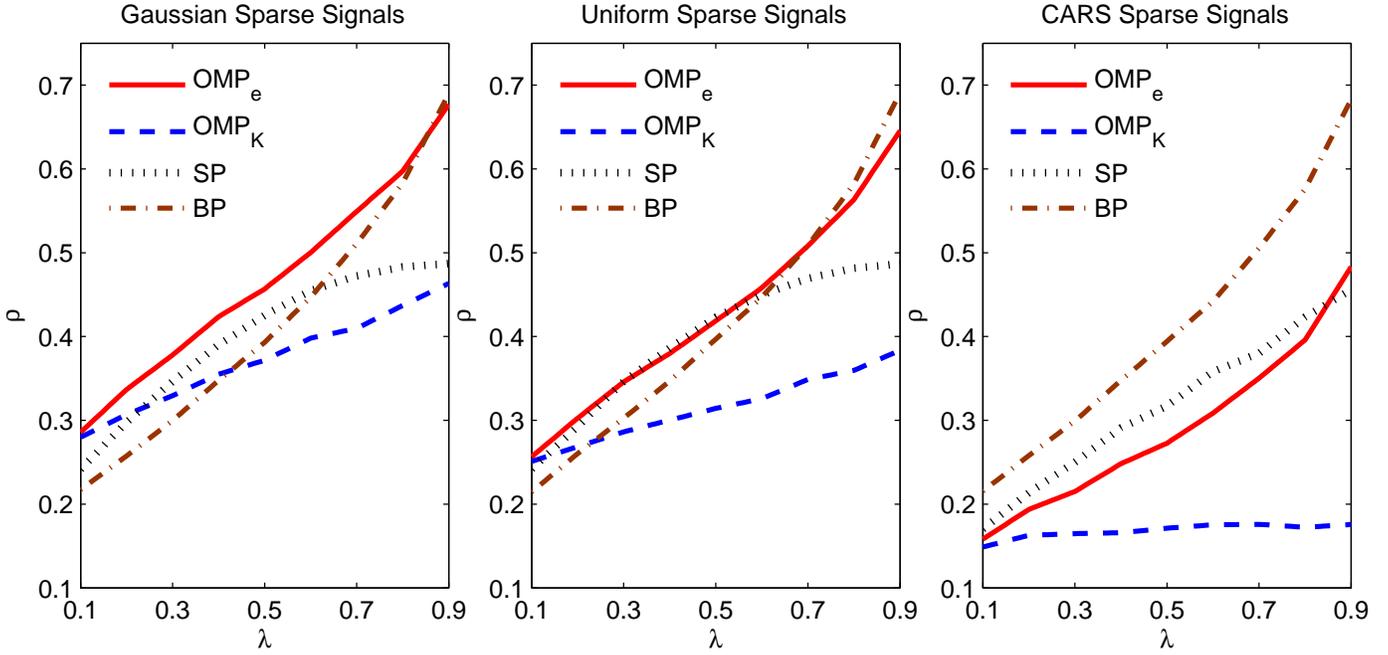}
\caption{Empirical phase transitions of OMP$_e$, OMP$_K$, BP and SP for the recovery of Gaussian, uniform and CARS sparse signals from noise-free observations. The entries of the observation matrices are selected as i.i.d. Gaussian random variables. The results are obtained over 200 trials. The axes labels are defined as $\rho = K/M$ and $\lambda = M/N$ where $N=250$.}
\label{fig:PT}
\end{center}
\end{figure*}

In this section, we compare the empirical recovery performances of OMP$_e$ and OMP$_K$ with basis pursuit (BP) \cite{Chen:BP} and SP via phase transitions.
We run the simulations for three different nonzero element distributions. The nonzero elements of the 'so-called' Gaussian sparse signals are drawn from the standard Gaussian distribution, while those of the uniform sparse signals are distributed uniformly in $[-1,1]$. The last ensemble involved is the Constant Amplitude Random Sign (CARS) sparse signals (following the definition in \cite{Maleki:TST}) where the nonzero elements have unit magnitude with random signs.
For OMP$_e$, $\varepsilon = 10^{-6}$ and the maximum allowable number of iterations is set to $M$. The exact recovery condition for $\mathbf{x}$ is specified as $\|\mathbf{x}-\mathbf{\tilde{x}}\|_2 \leq 10^{-2}\|\mathbf{x}\|_2$,\footnote{This choice is taken from \cite{Maleki:TST} to ensure compatibility of the computed phase transitions.} where $\mathbf{\tilde{x}}$ is the reconstructed sparse vector.

We compute the empirical phase transitions in order to provide an extensive evaluation over a wide range of $K$ and $M$. Let's define normalized measure for the number of observations as $\lambda = M/N$ and for sparsity level as $\rho = K/M$. We keep $N=250$ fixed, and alter $M$ and $K$ to sample the $\{\lambda, \rho\}$ space for $\lambda \in [0.1, 0.9]$ and $\rho \in [0,1]$. We randomly generate 200 sparse instances for each $\{\lambda,\rho\}$ tuple. Next, we draw a random Gaussian observation matrix for each test instance and run each algorithm to recover $\mathbf{x}$. After recovery of all samples, we compute the phase transitions by the methodology described in \cite{Maleki:TST}. This methodology uses a generalized linear model with logistic link to describe the exact recovery curve over $\rho$ for each $\lambda$. Then, the phase transition curve is finally given by combining the $\rho$ values which provide $50\%$ exact recovery rate for each $\lambda$.

\figurename~\ref{fig:PT} depicts the phase transition curves of OMP$_e$, OMP$_K$, BP and SP for the Gaussian, uniform and CARS sparse signals. OMP$_e$ yields clearly better phase transitions than OMP$_K$ does for all distributions as we intuitively expect.
On the other hand, the recovery performance of OMP highly depends on the coefficient distribution, while BP is robust to it, and SP shows less variation than OMP does. At one end stands the Gaussian sparse signals, where OMP$_e$ outperforms BP and SP. For the uniform sparse signals, OMP$_e$ might also be considered as the most optimal algorithm among the candidates over the whole $\lambda$ range. In contradiction, the performance of OMP degrades severely for the CARS ensemble, which is indeed referred to as the most challenging case for the greedy algorithms \cite{Dai:SP, Maleki:TST}.

These results clearly indicate the dependency of the OMP recovery performance on the coefficient distribution. When the nonzero values cover a wide range, such as for the Gaussian distribution, the performance of OMP is boosted. In contrast, nonzero values of equal magnitude constitute the most difficult recovery problem for OMP.
In fact, this dependency can be better explained by some basic analytical observations on $\|\mathbf{\Phi}_{T}^* \mathbf{y}\|_{\infty}$.
Assuming that the columns of $\mathbf{\Phi}$ are normalized, we can write the upper bound on $\|\mathbf{\Phi}_{T}^* \mathbf{y}\|_{\infty}$ as
\begin{IEEEeqnarray}[]{rcl}
\label{eq:chp3_bin}
\|  \mathbf{\Phi}_{T}^* \mathbf{y} \|_{\infty}  &{}={}&  \max_{t \in T} | \phi_{t}^* \Phi_{T} \mathbf{x}_{T} | \nonumber \\
&{}={}&  \max_{t \in T} | \phi_{t}^* \Phi_{T-t} \mathbf{x}_{T-t} + \phi_{t}^* \phi_{t} \mathbf{x}_{t} | \nonumber \\
&{}\leq{}& \max_{t \in T} | \phi_{t}^* \Phi_{T-t} \mathbf{x}_{T-t} | + | \phi_{t}^* \phi_{t} \mathbf{x}_{t} | \nonumber \\
&{}\leq{}& \max_{t \in T} \delta_K \|\mathbf{x}_{T-t} \|_2 + |\mathbf{x}_{t}|.
\end{IEEEeqnarray}
First, we do not force any restrictions on the nonzero values of $\mathbf{x}$. The Gaussian sparse signals can be seen an example of this case.
For simplicity, let us set $a = \|\mathbf{x}_{T-t} \|_2$.
Clearly, $0\leq a\leq\|\mathbf{x}\|_2$ in this setting. Hence, the upper bound on $\|\mathbf{\Phi}_{T}^* \mathbf{y}\|_{\infty}$ is given by
\begin{equation} \label{eq:chp3_bin1}
    \max\limits_{0\leq a\leq\|\mathbf{x}\|_2} a\delta_K + \sqrt{\|\mathbf{x}\|_2^2-a^2}.
\end{equation}
We simply take the derivative of (\ref{eq:chp3_bin1}) with respect to $a$, and set it equal to zero:
\begin{equation} \label{eq:chp3_bin2}
    \delta_K  - \frac{a}{\sqrt{\|\mathbf{x}\|_2^2-a^2}} = 0.
\end{equation}
Then, the $a$ value that maximizes (\ref{eq:chp3_bin1}) is found as
\begin{equation} \label{eq:chp3_bin3}
     a = \frac{\delta_K\|\mathbf{x}\|_2}{\sqrt{1+\delta_K^2}}.
\end{equation}
Replacing this into (\ref{eq:chp3_bin1}), we obtain
\begin{equation} \label{eq:chp3_bin4}
 \delta_K \|\mathbf{x}_{T-t} \|_2 + |\mathbf{x}_{t}| \leq \sqrt{1+{\delta_K}^2} \|\mathbf{x}\|_2.
\end{equation}
Consequently, the upper bound on $\|\mathbf{\Phi}_{T}^* \mathbf{y}\|_{\infty}$ is obtained as
\begin{equation} \label{eq:chp3_bin5}
\|  \mathbf{\Phi}_{T}^* \mathbf{y} \|_{\infty} \leq \sqrt{1+{\delta_K}^2} \|\mathbf{x}\|_2
\end{equation}
when there are no restrictions on the nonzero values of $\mathbf{x}$. Note that this upper bound defines the range which the values of the correlation vector at correct indices span during the first iteration.

\begin{figure*}[!t]
\begin{center}
\includegraphics[width=\linewidth]{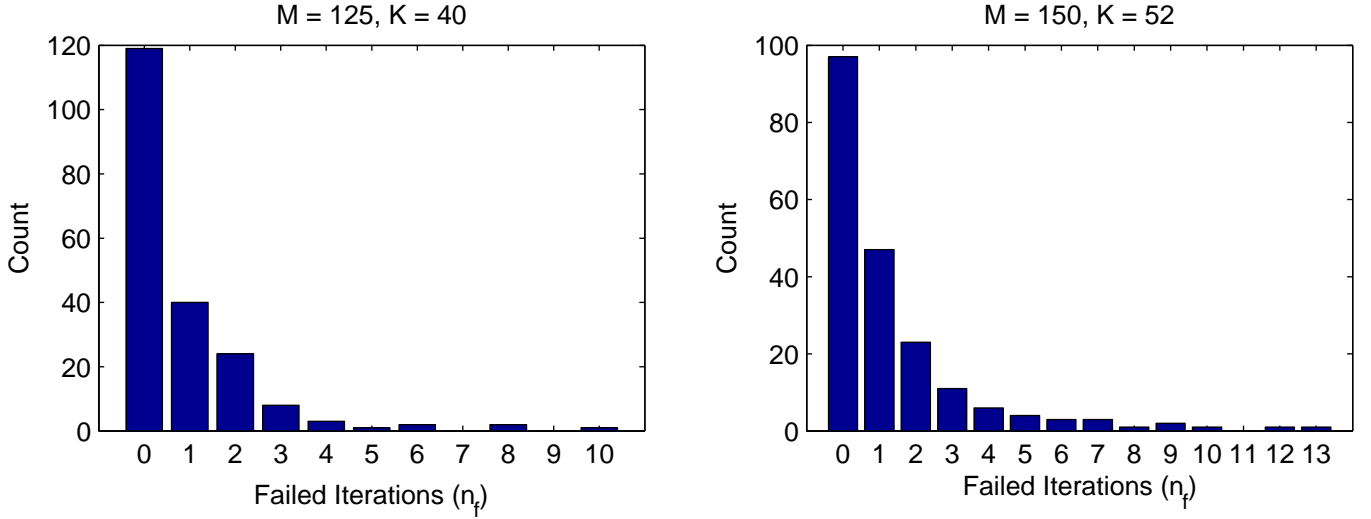}
\caption{Histograms of failed OMP$_e$ iterations ($n_f$) over 200 perfectly recovered Gaussian sparse vectors. $n_f = 0$ corresponds to the samples which are successfully recovered by OMP$_K$. OMP$_K$ perfectly recovers 119 out of 200 samples when $M = 125$, $N = 40$ and 97 out of 200 samples for $M = 150$, $N = 52$. OMP$_e$ recovers all samples perfectly in both cases. The number of failed OMP$_e$ iterations do not exceed $K/4$ for both cases.}
\label{fig:Hist}
\end{center}
\end{figure*}

Now, let's consider the CARS case, where $|\mathbf{x}_{t}|=1$ and $\|\mathbf{x}_{T-t} \|_2=\sqrt{K-1}$ for every $t \in T$. In this case, the upper bound on $\|\mathbf{\Phi}_{T}^* \mathbf{y}\|_{\infty}$ is given by
\begin{equation}
\label{eq:chp3_PT1}
| \mathbf{\Phi}_{T}^* \mathbf{y} |_{\infty} \leq 1 + \delta_K\sqrt{K-1}.
\end{equation}
The upper bound in (\ref{eq:chp3_PT1}) is obviously much smaller than the one in (\ref{eq:chp3_bin5}) in practice. (In order to compare them, fix the energy of $\mathbf{x}$, i.e., replace $\|\mathbf{x}\|_2 = \sqrt{K}$ into (\ref{eq:chp3_bin5}).)
This constitutes no problems if Theorem~\ref{Thrm_Wang} is satisfied. Consider, however, that Theorem~\ref{Thrm_Wang} fails:
In that case, the elements of $\mathbf{\Phi}^*\mathbf{y}$ at indices out of $T$ are more likely to exceed $|\mathbf{\Phi}_{T}^* \mathbf{y} |_{\infty}$ if $\mathbf{x}$ is a CARS sparse signal since $\|\mathbf{\Phi}_{T}^* \mathbf{y}\|_{\infty}$ is typically smaller for this kind of signals. Hence, the probability of failure at the first iteration becomes higher for the CARS sparse signals\footnote{Note that, though we skip it here, a similar analogy might be carried out to the following iterations as well.}. In other words, the maximum element of the correlation vector is less likely to be in the correct support for the CARS sparse signals, i.e., the correlation maximization step fails with higher probability.
As a result of this, it is natural that the failure rates of OMP-type algorithms increase when the range which is spanned by the absolute values of the nonzero elements of the underlying sparse signals decreases. The CARS signals have the smallest range of span, hence the worst performance of OMP-type algorithms naturally appears for these signals.
Note that this behaviour can be expected in common for all algorithms which employ a similar correlation maximization step. For example, Figure~\ref{fig:PT} indicates that the performance of SP, which employs a similar correlation maximization step, also decreases for sparse signals with constant amplitude nonzero elements.

\subsection{Empirical Success and Failure Rates on OMP$_e$ Iterations}
\label{Sec:Results_Hist}

Theorem~\ref{Thrm_OMP2} is based on the assumption $n_f < \lceil K/2 \rceil$, which leads to the other constraints on $K$ and $n_c$, i.e. $K \geq 25$ and (\ref{Eq:n_cBound2}).
Hence, satisfying the limit on the number of failed iterations is critical for Theorem~\ref{Thrm_OMP2}. On the other hand, the bound $n_f < \lceil K/2 \rceil$ may also be loose for many practical examples, making these constraints too restrictive in practice. Therefore, it is worthwhile to investigate the number of failed iterations after the termination in order to validate these constraints.

For this purpose, we choose two examples from the tests, and depict the histograms of $n_f$ after the successful termination of OMP$_e$.
The successful termination is important here as OMP$_e$ may run until it reaches the maximum number of iterations ($M$) in case of a failure, which makes the resultant histograms noninformative.
Therefore, we consider two cases where OMP$_e$ perfectly recovers all instances, while OMP$_K$ cannot, namely namely $M=125$, $K=40$ ($\lambda = 0.5$, $\rho=0.32$) and $M=150$, $K=52$ ($\lambda = 0.6$, $\rho=0.347$).
The histograms of failed iterations are depicted in Figure~\ref{fig:Hist}.
OMP$_K$ can only recover 119 out of 200 instances perfectly for the first case, and 97 for the latter.
For these instances, OMP$_e$ also provides perfect recovery with no failed iterations, hence these correspond to the region $n_f=0$ in the plots.
On the other hand, OMP$_e$ takes a number of wrong steps before finally finding the correct solution of the recovery problems where OMP$_K$ fails.
We observe that the number of these steps is smaller than the upper bound $\lceil K/2 \rceil - 1$. Actually, in both tests OMP$_e$ never takes more than $K/4$ wrong steps. Hence, the assumption $n_f < \lceil K/2 \rceil$ turns out to be empirically loose at least for these two cases.

\section{Conclusion}

In this manuscript, we have discussed theoretical and empirical analyses of the OMP recovery from noise-free observations with the termination criterion based on the residual power. This type of termination criterion presents a more suitable objective than setting the number of iterations equal to $K$ when the aim is finding an exact $K$-sparse representation, rather than obtaining the best $K$-sparse approximation.

The theoretical analyses in Section~\ref{Sec:Analysis} state an online recovery condition for OMP$_e$ based on the number of correct and false indices in the support estimate of an intermediate iteration. Though we cannot cast this condition into exact recovery guarantees for all $K$-sparse signals due to the lack of a proof for the existence of such support estimates, we still state that it may be satisfied online if $n_c$ and $n_f$ satisfy some bounds where OMP$_K$ recovery already fails.

On the other hand, as discussed in \ref{sec:RecentOMP}, the state-of-the-art more than $K$-step guarantees \cite{Zhang:OMP_RIP,Foucart:WOMP,Wang:Imp_OMP} are mostly impractical for many applications since they require a large number of iterations.
In contrast, according to Theorem~\ref{Thrm_OMP2}, the conditions presented in this letter may be imposed to provide online guarantees for recovery within $\frac{3}{2}K$ iterations. This is well below the number of iterations required for the state-of-the-art exact recovery guarantees of \cite{Zhang:OMP_RIP}, \cite{Foucart:WOMP}, and \cite{Wang:Imp_OMP}.

We have also demonstrated the recovery performance of OMP$_K$ and OMP$_e$ via simulations involving sparse signals with different nonzero coefficient distributions.
The phase transitions presented in Section~\ref{Sec:Results_PT} reveal that OMP$_e$ is capable of providing better recovery rates than BP and SP when the nonzero elements follow the Gaussian or uniform distributions.
Finally, we have presented histograms of the number of failed iterations in order to test the validity of the upper bound $n_f < \left\lceil \frac{K}{2} \right\rceil$. These histograms indicate that this upper bound is not only valid, but also loose in practice.

Regarding generalization of the developed online conditions as exact recovery guarantees for all $K$-sparse signals, future work may be conducted on the existence of support estimates satisfying the necessary conditions. Moreover, these conditions may be further improved by incorporating a tighter bound on either $\frac{\delta_{K+n_f+1}}{\delta_{K+1}}$ or $n_f$ as future work. To conclude, we believe that these findings will provide a basis for further improvements in the theoretical analyses of OMP and its variants.

\bibliographystyle{IEEEbib}
\bibliography{IEEEAbrv,AStarOMP}
%

\end{document}